\documentclass[12pt]{article}
\usepackage{amsmath, amsthm, amscd, amsfonts, amssymb, graphicx, color}
\usepackage[bookmarksnumbered, colorlinks, plainpages]{hyperref}
\hypersetup{colorlinks=true, linkcolor=red, anchorcolor=green, citecolor=green, urlcolor=black, filecolor=black, pdftoolbar=true}
\theoremstyle{definition}

\theoremstyle{remark}

\numberwithin{equation}{section}

\begin{document}

	\begin{center}
 \textbf{ Relativistic isotropic stellar model with Durgapal-V metric in $f({R},  {T})$ gravity} \\

		\bigskip
	Nayan Sarkar,$^{1,a}$ 	Susmita Sarkar,$^{2,b}$  Banashree Sen $^{3,c}$  Moumita Sarkar,$^{4,d}$ and  Farook Rahaman$^{4,*}$     \\

	$^{1}$ Department of Mathematics, Karimpur Pannadevi College, Karimpur-741152, Nadia, West Bengal, India

	$^{2}$ Department of Applied Science and Humanities, Haldia Institute of Technology, Haldia-721606, West Bengal, India

$^{3}$ Department of Commerce and Management, St. Xavier's University, Kolkata-700160, India

$^{4}$ Department of Mathematics, Jadavpur University, Kolkata-700 032, India

	\footnote{Email addresses:  $^{a}$\url{nayan.mathju@gmail.com}, $^{b}$\url{susmita.mathju@gmail.com}, $^{c}$\url{banashreesen7@gmail.com} , $^{d}$\url{moumita.sarkar1594@gmail.com} , $^{*}$\url{rahaman@associates.iucaa.in}, }

	\end{center}
	
	\bigskip

	\begin{abstract}
The main aim of this paper is to obtain a completely new relativistic non-singular model for static, spherically symmetric isotropic celestial compact stars in the  $f(R, T)$ gravity scenario. In this regard, we have considered the isotropic Durgapal-V metric {\it ansatz} \cite{dur82} to find the solutions of Einstein's field equations in the framework of $f(R, T)$ gravity. The obtained solutions are analyzed graphically for the compact star {\it Cen X}-3 with mass $M$ = $ 1.49 \pm 0.08 ~ M_\odot$ and radius $R$ = 9.178 $\pm$ 0.13 $km$ \cite{ml11} and numerically for ten well-known different compact stars along with  {\it Cen X}-3 corresponding to the different values of coupling constant $\chi$.  The reported solutions are singularity-free at the center of the stars, physically well-behaved, and hold the physically stable matter configurations by satisfying all the energy conditions and EoS parameter $\omega(r) \in $ (0, 1), causality condition, adiabatic index $\Gamma(r) > 4/3$. We have also discussed hydrostatic equilibrium through the modified TOV equation to ensure the equilibrium position of the solutions representing matter distributions. Considering the several values of $\chi$ we have examined the impact of this parameter on the proposed solutions that help to make a fruitful comparison of modified $f(R, T)$ gravity to the standard general relativity,  and interestingly, we have found that the modified $f(R, T)$ gravity holds long-term stable compact objects than the standard Einstein gravity. All the graphical and numerical results ensure that our reported model is under the physically admissible regime that indicates the acceptability of the model.   \\

\end{abstract}

%\email{ }
%\affiliation{}

%\date{\today}

%\keywords{$f(R, T)$ gravity, Durgapal-V Metric, Compact Star, Isotropy. }% Use showkeys class option if keyword

%\begin{framed}
%\tableofcontents
%\end{framed}

\section{Introduction}\label{Sec1}

 The  Supernova Search Team\cite{sst,sst1}, the Supernova Cosmology Project\cite{scp,scp1,scp2}, the Wilkinson Microwave Anisotropy Probe (WMAP)\cite{wmap,wmap1,wmap2}, and the Sloan Digital Sky Survey (SDSS)\cite{sdss,sdss1} separately discovered that the present universe is expanding in an accelerated order. Later, this discovery encouraged the scientific community to find out the reason behind the accelerated expansion of the universe. This leads to the fact that the accelerated expansion of the universe is happening due to some kind of mysterious hidden energy with a huge amount of negative pressure, known as the Dark Energy (DE). To investigate the mystery of DE several researchers developed different modified theories of gravity of the standard general theory of relativity, like, unimodular gravity\cite{noj16,gar19}, $f(R)$ gravity\cite{noj07,noj11,fs09, tp10, sc10}, $f(R, G)$ gravity\cite{ad11,ka14},$f(G, T)$ gravity\cite{ms16,mz18},  $f(Q, T)$ gravity \cite{xu19,aro20}, teleparallel gravity\cite{ra13,sb22}, and so on.  The $f(R)$  gravity is introduced by replacing the Ricci scalar $R$ with an arbitrary generalized function of  $R$ in the standard Einstein Hilbert action and it can explain the late time cosmic acceleration, unified inflation with dark energy and galactic dynamics of massive test particles\cite{snn07,sn008,snn11,gc08,ee11}. Harko et al.\cite{har11}  introduced a more generalized theory of gravity than the f(R) gravity by considering the gravitational Lagrangian of the standard Einstein Hilbert action as the arbitrary function $f(R, T)$ where $R$ is the Ricci scalar and $T$ is the trace of the energy-momentum tensor. After the origination of $f(R, T)$ gravity, researchers studied the behavior of perfect fluid and massless scalar field for homogeneous and anisotropic Bianchi type I universe model, and the energy conditions in the framework of $f(R, T)$ gravity\cite { ms012, Mss13}. The $f(R, T)$ gravity also explained several cosmological applications\cite{ph14,ph15,ph16,cp14,hs14,dr13,pk15}. From the last decade, the $f(R, T)$ theory of gravity has also become interesting at the astrophysical level, and several researchers successfully investigated different applications of this gravity\cite{MS14,IN15,IN15A,IN15B,AA15,MZ15,MZ16,AD17,DD18,MS18,MS19,sk20}. Furthermore,  Das et al.\cite{AD16} studied the interior solutions of a compact star admitting conformal motion in the framework of $f(R, T)$ gravity. Moraes et al.\cite{PH16} used the $f(R, T)$ gravity framework to study the hydrostatic equilibrium configuration of neutron stars and strange stars. Waheed et al.\cite{sw20}, Zubair et al.\cite{mz21}, and Sarkar et al.\cite{ss22} separately developed different new models for the compact stars satisfying the Karmarkar and Pandey-Sharma conditions in $f(R, T)$ gravity. The relativistic stellar structures with the variable cosmological constant are also studied in the $f(R, T)$ theory of gravity\cite{mi20}. Pretel et al.\cite{jm21} studied the equilibrium and stability of the celestial compact objects by assuming a polytropic equation of state, Lobato et al.\cite{rl20} studied the neutron stars by considering realistic equations of state, and Shamir et al.\cite{ms018} studied the spherically symmetric anisotropic by considering the {\it MIT} bag model equation of state in the framework of $f(R, T)$ gravity.  Recently, Bhar et al.\cite{pb22} studied the isotropic Buchdahl relativistic fluid sphere in the ground of $f(R, T)$ gravity.

Several astrophysicists have continuously tried to introduce perfect fluid models for superdense celestial matter configurations since the discovery of Einstein's field equations. In the year 1916, Schwarzchild \cite{ks16} first found the exact solution of the Einstein field for the interior of hydrostatic equilibrium perfect matter configuration, this significant study provokes researchers to find the exact solutions satisfying all the necessary conditions for physical acceptance\cite{hk88}. In the year 1939, Tolman\cite{rc39}  developed a new method to solve Einstein's field equations for static spheres of perfect fluid spheres, and Adler\cite{rj74} also solved the field equations for the interior of a static perfect fluid sphere in the context of standard general relativity in a different technique. Matese et al.\cite{jj80} developed a new formalism to the static spherical symmetry perfect fluid sphere in Schwarzschild coordinates. Later, Rahaman and Visser\cite{sh02} introduced an explicit metric for the static spherically symmetric perfect-fluid spacetime, and Lake\cite{kl03} provided an algorithm based on the choice of a single monotone function for presenting all regular static spherically symmetric perfect-fluid solutions of Einstein's field equations. Furthermore, Pant et al.\cite{np10} presented the spherically symmetric regular solutions for relativistic perfect fluid spheres, and Prasad et al.\cite{ak21} studied the charged isotropic compact stars model with Buchdahl metric in general relativity. In this context, the different modified gravity also becomes the suitable framework to study the isotropic compact stars.  Abbas\cite{ga18} studied a completely new solution for an isotropic matter distribution in the framework of Rastall gravity. Hansraj\cite{sh20} studied the isotropic matter distributions in the framework of  4D Einstein Gauss-Bonnet gravity and Nashed\cite{gg23} studied the isotropic matter distributions in 4D Einstein Gauss-Bonnet gravity coupled with a scalar field.

In the year 1982, Durgapal\cite{dur82} introduced a class of new exact solutions for static spherically symmetric isotropic matter distributions by considering a simple relation of the metric potential function, this pioneering work of Durgapal\cite{dur82} creates a new dimension to the study of isotropic compact stars. Fuloria et al.\cite{pf12} and Mehta et al.\cite{rn13} studied the well-behaved charge analogue of the Durgapal solution.  Contreras et al.\cite{ec22} studied the uncharged and charged like-Durgapal models by using the vanishing complexity factor. Murad et al.\cite{mh13} and Maurya et al.\cite{sm11} presented the interior solutions of the Einstein-Maxwell field equations for a static spherically symmetric charged perfect fluid in the framework of general relativity with Durapal metric.  Islam et al.\cite{ri19} studied the strange stars in the Durgapal-IV spacetime and the quintessence compact star is also studied in the Durgapal spacetime\cite{ge20}. Very recently,  Rej\cite{pr23} studied the uncharged isotropic compact stars model in the bigravity with the Durgapal-IV metric, also, Rej et al.\cite{pr23a} studied the isotropic Durgapal-IV relativistic fluid sphere in the ground of $f(R, T)$ gravity. In this present work, our aim is to obtain the regular and physically well-behaved uncharged solutions of the Einstein field equations by considering the Durgapal-V metric potentials\cite{dur82} in the framework of $f(R, T)$ = $R + 2\chi T$ gravity. In this regard, we shall compare the proposed solutions of $f(R, T)$ gravity with the standard general relativity. The present solutions are analyzed with the help of the well-known compact star {\it Cen X}-3. The compact star {\it Cen X}-3 is the most luminous X-ray pulsar in our galaxy, and it was discovered with the help of a rocket-borne detector in the year 1967\cite{gc67}. Later, Giacconi et al.\cite{rg71} and  Schreier et al. \cite{es72} found the binary and pulsar nature of the  {\it Cen X}-3 from the satellite observations.

The present article is designed as follows: The Einstein field equations for static and spherically symmetric uncharged isotropic matter distributions are mentioned in Sec. \ref{Sec2}.  We have considered the Durgapal-V metric potential functions to solve the field equations in Sec. \ref{Sec3}. We have analyzed the matching of the external solution with our internal solutions at the surface of the compact star in Sec. \ref{Sec4}.  The relevant physical attributes of the proposed model are discussed in Sec. \ref{Sec5}. Sec. \ref{Sec6} is dedicated to the analysis of the equilibrium situation of the model via the TOV equation. We have analyzed the stability of the model in Sec. \ref{Sec7}. Sec. \ref{Sec8} deals with the moment of inertia of the system.  Finally, the results and conclusion of the present model are presented in Sec. \ref{Sec9}.

\section{Einstein's  field equations in $f(R,T)$ gravity }\label{Sec2}

 The fundamental pillar of the theory of  $f(R,T)$ gravity is based on the utilization of an algebraic general functional form of the Ricci scalar $R$ and the trace of energy-momentum tensor $T$ in the standard Einstein-Hilbert action. Harko {\it et al.}\cite{har11} described the Einstein-Hilbert action for the theory of $f(R,T)$ gravity in the following form
\begin{equation}
S= \frac{1}{16\pi}\int f(R,T)\sqrt{-g}~d^4x +\int  {L}_m \sqrt{-g}~d^4x.\label{S2}
\end{equation}
 where $f(R, T)$, as mentioned earlier, is a general function of $R$ with $T$, $g$ is the determinant of the metric tensor $g_{\mu\nu}$ and  $ {L}_m$ stands for the matter Lagrangian density related to the energy-momentum tensor $T_{\mu\nu}$. Now, the energy-momentum tensor of the matter distribution can be written as\cite{ld13}
\begin{equation}
T_{\mu\nu}=-\frac{2}{\sqrt{-g}}\frac{\delta(\sqrt{-g} {L}_m)}{\delta g^{\mu\nu}}. \label{T1}
\end{equation}
 Here, the trace $T = g^{\mu\nu}T_{\mu\nu}$. Let us suppose that the Lagrangian density $ {L}_m$ acts as a function of $g_{\mu\nu}$ not its derivatives\cite{har11}, then Eq. (\ref{T1}) reads as
 \begin{equation}
T_{\mu\nu}= {L}_m g_{\mu\nu}-2\frac{\partial  {L}_m}{\partial g^{\mu\nu}}. \label{T2}
\end{equation}

Now, the Einstein field equations in the background of $f(R, T)$ gravity with the Einstein-Hilbert action (\ref{S2})  can be written as
\begin{eqnarray}
 \bigtriangledown_\mu\bigtriangledown^\mu g_{\mu\nu} f_R(R,T)-\frac{1}{2}f(R,T)g_{\mu\nu}+(R_{\mu\nu}-\bigtriangledown_\mu \bigtriangledown_\nu)f_R(R,T)
=8\pi T_{\mu\nu}-f_T(R,T)(T_{\mu\nu}+\Theta_{\mu\nu}),\label{eq1}
\end{eqnarray}
where $ \bigtriangledown_\mu\bigtriangledown^\mu \equiv \frac{1}{\sqrt{-g}}\frac{\partial}{\partial x^\mu}(\sqrt{-g}g^{\mu\nu}\frac{\partial}{\partial x^\nu})$ is the D'Alembert operator, $f_R(R,T)=\frac{\partial f(R,T)}{\partial R}$, $f_T(R,T)=\frac{\partial f(R,T)}{\partial T}$, $R_{\mu\nu}$ is the Ricci scalar,  $\bigtriangledown_\mu$ stands for the covariant derivative associated with the Levi-Civita connection of $g_{\mu\nu}$ and $\Theta_{\mu\nu}=g^{\alpha\beta}\frac{\delta T_{\alpha\beta}}{\delta g^{\mu\nu}}$.

On applying the covariant derivative on Eq. (\ref{eq1}), one can get the following result\cite{har11,tk06}
\begin{eqnarray}
\bigtriangledown^\mu T_{\mu\nu}&=&\frac{f_T(R,T)}{8\pi-f_T(R,T)}\left\{\bigtriangledown^\mu \Theta_{\mu\nu} +(T_{\mu\nu}+\Theta_{\mu\nu})\bigtriangledown^\mu \ln f_T(R,T)\right\}.\label{aa}
\label{eq2}
\end{eqnarray}
The above equation (\ref{eq2}) shows that $\bigtriangledown^\mu T_{\mu\nu} \neq 0$ whenever $f_T(R,T) \neq 0 $ i.e. the energy-momentum tensor is not conserved as like Einstein gravity. Actually, the coupling between matter and curvature in $f(R, T)$ gravity creates a non-conserved stress-energy tensor, therefore, an extra force will be generated within the matter configuration that plays a crucial role in equilibrium. However, the energy-momentum tensor is conserved for $f(R, T)$ = $R$, and the corresponding field equations (\ref{eq1}) reduce for the Einstein gravity.
For the present model, we consider the following algebraic form of $f(R, T)$\cite{har11}
\begin{equation}
f(R,T)=R+2\chi T.\label{frt}
\end{equation}
where $ \chi $ is known as the coupling constant. For this choice of $f(R, T)$, the field equation (\ref{eq1}) will reduce into the Einstein gravity whenever $ \chi $ = 0.

Now, the line element to describe the interior of a  static and spherically symmetric stellar matter configuration in the Schwarzchild coordinate system ($t$, $r$, $\theta$, $\phi$) can be written as
\begin{equation}
ds^2 = e^{\nu(r)}dt^2 - e^{\lambda(r)}dr^2 - r^2(d\theta^2 + \sin^2{\theta}d\phi^2). \label{metric}
\end{equation}

where $e^{\nu(r)}$ and $e^{\lambda(r)}$ are called the metric potentials that are the functions of the radial coordinate $r$ only.

We consider that the matter composition within the sellar structure is perfect fluid, and hence, the corresponding energy-momentum can be written as

\begin{eqnarray}
T_{\mu\nu}&=&\{\rho(r)+P(r)\}u_\mu u_\nu-P(r)g_{\mu\nu}, \label{eq4}
\end{eqnarray}
where $\rho(r)$ and $P (r)$ are the energy density and isotropic pressure of the matter configuration in the $f(R, T)$ gravity, respectively, and $u_\nu$ is the four-velocity satisfying $u_\mu u^\mu=-1$ and $u_\nu \bigtriangledown^\mu u_\mu=0$. Here, we have taken the lagrangian defined as $ {L}_m$ = -$P(r)$ \cite{har11}, and therefore, $\Theta_{\mu\nu} = -2T_{\mu\nu}-P(r)g_{\mu\nu}$.

Now, the Einstein field equations (\ref{eq1}) for considered form of $f(R, T)$, given in Eq. (\ref{frt}) take the following form
\begin{eqnarray}
T_{\mu\nu}^{eff}&=& T_{\mu\nu} +\frac{\chi}{8\pi}Tg_{\mu\nu}+\frac{\chi}{4\pi}\left(T_{\mu\nu}+P(r)g_{\mu\nu}\right), \label{eq6}
\end{eqnarray}

Therefore, the  field equations (\ref{eq6}) for the spacetime (\ref{metric}) in $f(R, T)$ gravity  read as

\begin{eqnarray}
\rho_{eff}(r)&=&\frac{e^{-\lambda(r)}}{8\pi}\left\{\frac{\lambda^\prime(r)}{r}-\frac{1}{r^2}\right\}+\frac{1}{8\pi r^2}, \label{rhoeff}
\\
 P_{{eff}}(r)&=&\frac{e^{-\lambda(r)}}{8\pi}\left\{\frac{\nu^\prime(r)}{r}+\frac{1}{r^2}\right\}-\frac{1}{8\pi r^2}, \label{preff}
\\
P_{{eff}}(r)&=&\frac{e^{-\lambda(r)}}{32\pi}\left\{2\nu^{\prime\prime}(r)+\nu^{\prime 2}(r)+\frac{2(\nu^{\prime}(r)-\lambda^{\prime}(r))}{r}-\nu^{\prime}(r)\lambda^{\prime}(r)\right\},\label{pteff}
\end{eqnarray}

where $'$ denotes the derivative with respect to the radial coordinate $r$, $\rho_{eff}(r)$ and $P_{{eff}}(r)$ are the effective energy density and effective isotropic pressure that are related to the energy density $\rho(r)$ and isotropic pressure  $P(r)$ of $f(R, T)$ gravity as

\begin{eqnarray}
\rho_{eff}(r)&=&\rho(r)+\frac{\chi}{8\pi}\{3\rho(r)-P(r)\},\label{rho}
\\
P_{r_{eff}}(r)&=&P(r)-\frac{\chi}{8\pi}\{\rho(r)-3P(r)\}.\label{P}
\end{eqnarray}

%%%%%%%%%%%%%%%%%%%%%%%%%%%%%%%%%%%%%%%%%%%%%%%%%%%%%%%%%%%%%%%%%%%%%
\begin{figure}[!htbp]
\begin{center}
\begin{tabular}{rl}
\includegraphics[width=8.5cm]{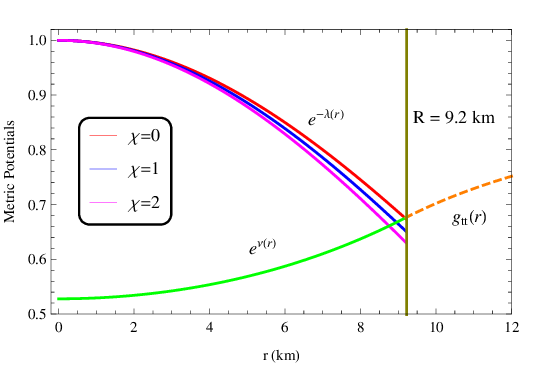}
\includegraphics[width=8.56cm]{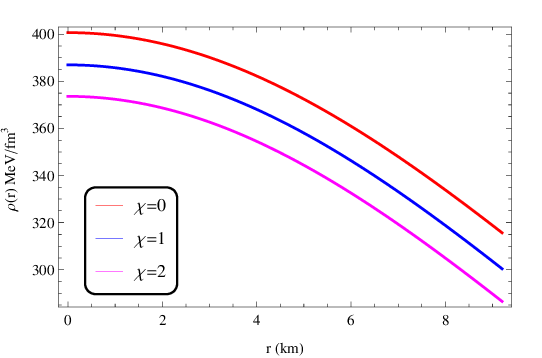}
\\
\end{tabular}
\end{center}
\caption{  Profiles of metric potentials (Left) and energy density (Right) against the radial coordinate $r$ for the compact star {\it Cun X} -3 corresponding to the numerical values of constants given in Table-{\ref{tab1}}.}\label{fig1}
\end{figure}
%%%%%%%%%%%%%%%%%%%%%%%%%%%%%%%%%%%%%%%%%%%%%%%%%%%%%%%%%%

From equations (\ref{rhoeff})-(\ref{pteff}) we get
\begin{eqnarray}\label{gg}
    \frac{\nu'(r)}{2}[\rho(r)+P(r)]+\frac{dP(r)}{dr}=\frac{\chi}{8\pi+2\chi}\left[P'(r)-\rho'(r)\right].
\end{eqnarray}
Again, one can see from Eq. (\ref{gg}) that the conservation equation in Einstein's gravity is obtained for $\chi = 0$. Further, the  EoS parameter is defined as $\omega(r) = P(r)/\rho(r)$. To generate the present model we shall adopt the well-known isotropic Durgapal-V metric potentials in the next section.

\section{ Exact Solutions with Durgapal-V metric  }\label{Sec3}

 To introduce a physically well-behaved model for static and spherically symmetric celestial isotropic matter configurations we consider the Durgapal-V space-time metric {\it ansatz}\cite{dur82}, given by

\begin{eqnarray}
e^{\lambda(r)} &=& \left[\frac{A C r^2 }{\left(1+C r^2\right)^3(1+6 C r^2)^{1/3}}+\frac{1}{\left(1+C r^2\right)^3}\left\{1-\frac{C r^2}{112}  \left(8 C^2 r^4+54 C r^2+309\right)\right\}\right]^{-1},~~~~~~~\label{elam}
\\
e^{\nu(r)} &=& B(1+Cr^2)^5.\label{enu}
\end{eqnarray}
where $A$ and $B$ are dimensionless constants and $C$ is a constant with dimension of ${\it length^{-2} }$. Later, we will determine the values of $A$ and $B$ in terms of $C$ from the matching conditions.
It should be noted that the physical and geometric singularities need to be avoided in the study of compact star modeling. In this regard,  we examine the exact characteristics of metric potentials $e^{\nu(r)}$ and $e^{\lambda(r)}$ for the confirmation of non-singularity within the compact star. The derivatives of the metric potentials are obtained as
\pagebreak

\[\frac{de^{\lambda(r)}}{dr} =\]   \begin{equation}  \frac{224 C r (1 + C r^2)^2 \left[112 A (14 C^2 r^4 - 1 - 2 C r^2) +
   15 (1 + 6 C r^2)^{\frac{1}{3}} (43 + 224 C r^2 - 206 C^2 r^4 - 12 C^3 r^6)\right]}{(1 + 6 C r^2)^{2/3} \left[(1 + 6 C r^2)^{1/3} (309 C r^2 + 54 C^2 r^4 + 8 C^3 r^6 - 112) - 112 A C r^2\right]^2},\end{equation}
\begin{eqnarray}
\frac{de^{\nu(r)}}{dr} &=& 10 BCr(1 + C r^2)^4.\label{eos}
\end{eqnarray}
 
 From the above equations, we can see $\frac{de^{\lambda(0)}}{dr}$ = $\frac{de^{\nu(0)}}{dr}$ = 0, and therefore, the considered metric potentials are singularity free at the center of the star. To see the exact behaviors, we demonstrate $e^{\nu(r)}$ and $e^{-\lambda(r)}$ for the compact star {\it Cen X}-1 in Fig. \ref{fig1} (Left) that shows that both are regular and positively finite inside the star. Moreover, $e^{\nu(r)}$ is increasing and $e^{-\lambda(r)}$ is decreasing in nature and they have matched together with the exterior Schwarzechild solutions at the surface corresponding to $\chi$ = 0 only, this happens because of independence of $e^{\nu(r)}$  on $\chi$.

%%%%%%%%%%%%%%%%%%%%%%%%%%%%%%%%%%%%%%%%%%%%%%%%%%%%%%%%%%%%%%%%%%%%%
\begin{figure}[!htbp]
\begin{center}
\begin{tabular}{rl}
\includegraphics[width=8.5cm]{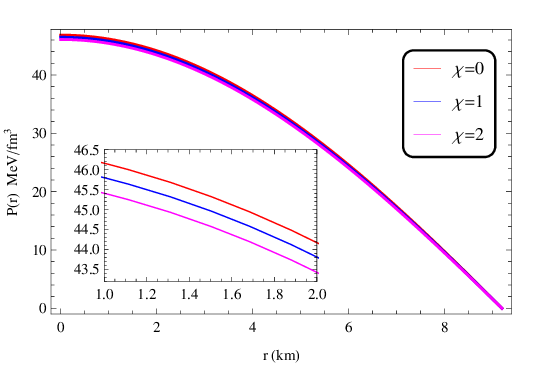}
\includegraphics[width=8.5cm]{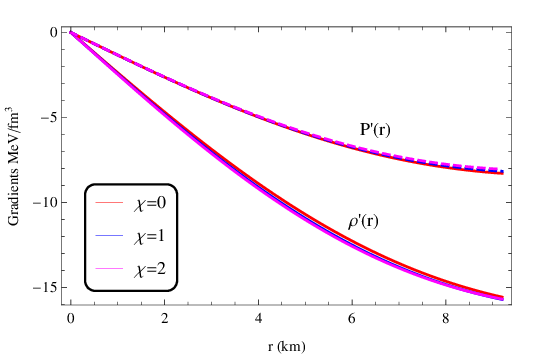}
\\
\end{tabular}
\end{center}
\caption{ Profiles of pressure (Left) and density gradient, pressure gradient (Right) against the radial coordinate $r$ for the compact star {\it Cen X}-3 corresponding to the numerical values of constants given in Table-{\ref{tab1}}.}\label{fig2}
\end{figure}
%%%%%%%%%%%%%%%%%%%%%%%%%%%%%%%%%%%%%%%%%%%%%%%%%%%%%%%%%%

On using the metric potentials (\ref{elam})-(\ref{enu}) we obtain the effective matter density and pressure from Eqs.(\ref{rhoeff})-(\ref{pteff}) as
\pagebreak
\[\rho^{eff}(r) =\]
\begin{equation}\frac{C \left[112A(22C^2r^4 -11Cr^2 -3) +
   15(1 + 6Cr^2)^{1/3} (129 + 775Cr^2 + 36C^2r^4 + 188C^3r^6 + 48 C^4r^8)\right]}{896 \pi (1 + C r^2)^4 (1 + 6 C r^2)^{4/3}},\label{rheff}\end{equation}
\begin{eqnarray}
P^{eff}(r) &=& \frac{C \left[112A(1 +11Cr^2) - 25(1+6Cr^2)^{1/3} (Cr^2(165+42Cr^2+8C^2r^4)-19)\right]}{896\pi(1+Cr^2)^4(1+6Cr^2)^{1/3}}.~~~~~~~~~~~~~\label{peff}
\end{eqnarray}

The above effective results determine the exact expressions of the energy density and pressure in $f(R, T)$ gravity from Eqs. (\ref{rho})-(\ref{P}) as

\begin{eqnarray}
\rho(r) &=&  \frac{C \left[112 A (\pi (44 C^2 r^4 - 22 C r^2 - 6) + (33 C^2 r^4 - 4 C r^2 -2) \chi) + 5\xi_1(r) (1 + 6 C r^2)^{4/3} \right]}{224 (1 + C r^2)^4 (1 + 6 C r^2)^{4/3} (8 \pi^2 + 6 \pi \chi + \chi^2)},~~~~~~\label{rh}
\\
P(r)  &=& \frac{C\left[112 A (2 \pi (1 + 17 C r^2 + 66 C^2 r^4) +
    5 C r^2 (2 + 11 C r^2) \chi) + 5 \xi_2(r) (1 + 6 C r^2)^{4/3} \right]}{224 (1 + C r^2)^4 (1 + 6 C r^2)^{4/3} (8 \pi^2 + 6 \pi \chi + \chi^2)},~~~~~~~\label{p}
\end{eqnarray}

where
\begin{eqnarray}
\xi_1(r) &=& 6 \pi (129 + C r^2 + 30 C^2 r^4 + 8 C^3 r^6) + (314 - 204 C r^2 + 15 C^2 r^4 + 8 C^3 r^6), \chi
     \\
 \xi_2(r) &=& 10 \pi (19 - 165 C r^2 - 42 C^2 r^4 - 8 C^3 r^6) +
  3 (56 - 206 C r^2 - 45 C^2 r^4 - 8 C^3 r^6) \chi.
\end{eqnarray}

 The physical parameters $\rho(r)$ and $P(r)$ are graphically demonstrated in Figs. (\ref{fig1}) (Right) and (\ref{fig2}) (Left), respectively, which show that they are non-singular and attain their maximum positive values at the center and thereafter monotonically positively decreasing towards the surface $r = R$ of the matter configuration to reach their minimum values with $\rho(R)\neq$ 0 and $P(R)$ = 0, ensuring that $\rho(r)$ and $P(r)$ are physically well-behaved and do not have any kind of singularity. Interestingly, we can also see the effect of the coupling constant $\chi$ on the matter density and pressure from Figs. (\ref{fig1}) (Right) and (\ref{fig2}) (Left), both are decreasing with increasing values of  $\chi \in$ [0, 2].

%%%%%%%%%%%%%%%%%%%%%%%%%%%%%%%%%%%%%%%%%%
\begin{figure}[!htbp]
\begin{center}
\begin{tabular}{rl}
\includegraphics[width=6.cm]{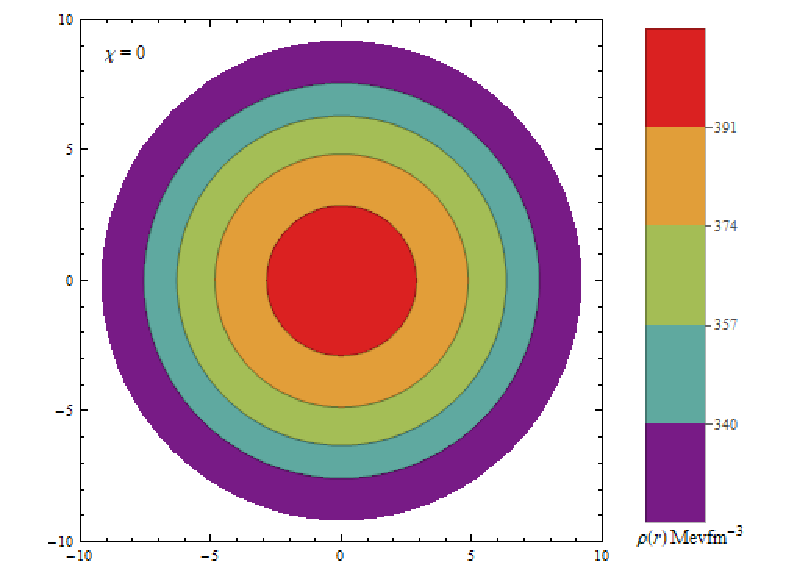}
\includegraphics[width=5.8cm]{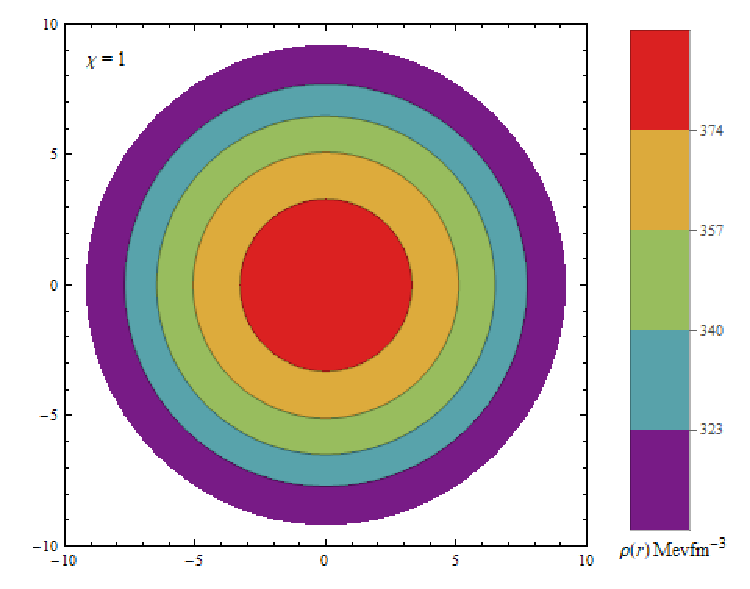}
\includegraphics[width=6.cm]{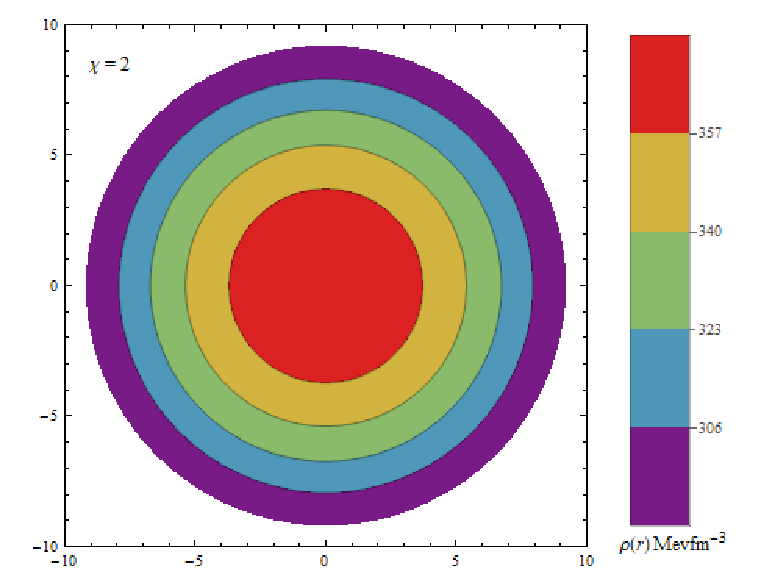}
\\
\end{tabular}
\begin{tabular}{rl}
\includegraphics[width=5.85cm]{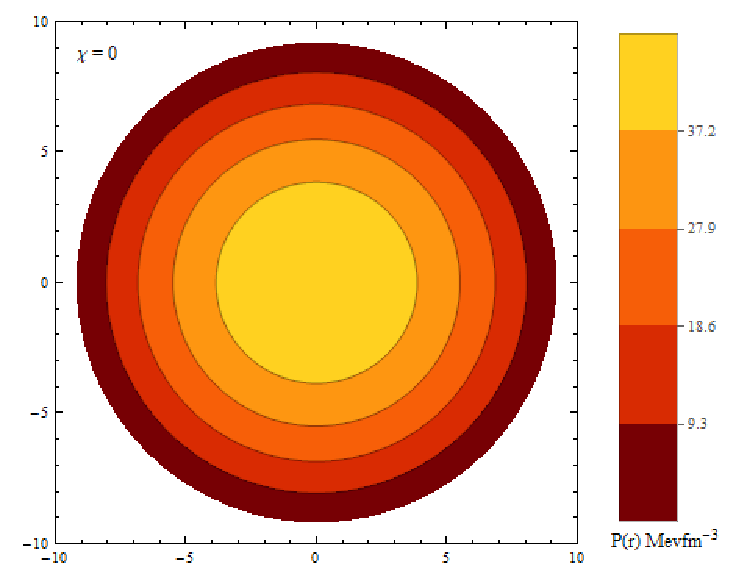}
\includegraphics[width=5.95cm]{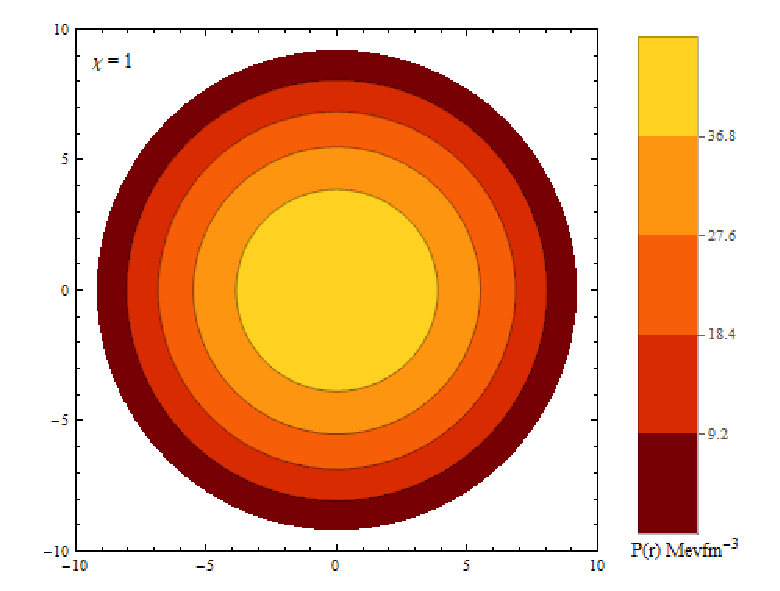}
\includegraphics[width=5.75cm]{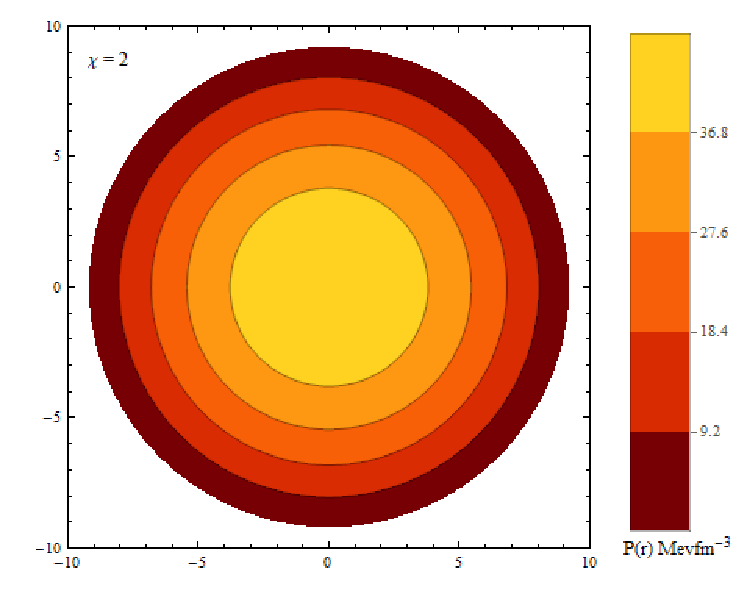}
\\
\end{tabular}
\end{center}
\caption{  Radially symmetric profiles of the energy density (Upper Panel) and  pressure (Lower Panel)  for the compact star {\it Cen X}-3 corresponding to the numerical values of constants given in Table-{\ref{tab1}}.}\label{fig3}
\end{figure}

%%%%%%%%%%%%%%%%%%%%%%%%%%%%%%%%%%%%%%%%%%%%%%%%%%%%

\section{ External solutions and  Matching with the  Internal solution  }\label{Sec4}

 The {\it Schwarzschild} exterior  solutions of the Einstein field equations can be written as
\begin{eqnarray}
ds_+^{2} &=& -g_{tt}(r)dt^{2}+g_{rr}(r)dr^{2} +r^{2}\left(d\theta^{2}+\sin^{2}\theta d\phi^{2} \right),\label{es}
\end{eqnarray}
where $g_{tt}(r) = g_{rr}^{-1}(r) = \left(1-\frac{2M}{r}\right)$ with $M$ as the mass of the matter configuration.

Now, Birkhoff's theorem says that the gravitational field of the outside of any isotropic or anisotropic spherically symmetric static celestial compact object is of Schwarzschild form. Therefore, for a physical model, the interior solutions need to be matched to the {\it Schwarzschild} exterior solutions at the surface $r = R > 2M$ of the fluid sphere, this matching helps to determine the values of some model-dependent constants. Our reported internal spacetime is

\[ds_-^{2} = -B(1+Cr^2)^5(r)dt^{2}\]\begin{eqnarray} &+& \left[\frac{A C r^2 }{\left(1+C r^2\right)^3(1+6 C r^2)^{1/3}}+\frac{1}{\left(1+C r^2\right)^3}\left\{1-\frac{C r^2}{112}  \left(8 C^2 r^4+54 C r^2+309\right)\right\}\right]^{-1}dr^{2}\nonumber
\\
&& +r^{2}\left(d\theta^{2}+\sin^{2}\theta d\phi^{2} \right).\label{is}
\end{eqnarray}

Therefore, for the matching between internal solutions (\ref{is}) and {\it Schwarzschild} external  solutions (\ref{es}) at the surface $r = R$ of the compact star we obtain the following results
\begin{eqnarray}
 && ~~~~~~~~~ B(1+CR^2)^5 = 1-\frac{2M}{R},~~~~~~~~~\label{mc}
 \\
 &&\frac{A C R^2}{\left(1+C R^2\right)^3(1+6 C R^2)^{1/3}} +\frac{1}{\left(1+C R^2\right)^3}\left\{1-\frac{C R^2}{112}  \left(8 C^2 R^4+54 C R^2+309\right)\right\} = 1-\frac{2M}{R}.~~~~~~~\label{mc,}
 \end{eqnarray}
 Also, as the  pressure becomes zero at the surface of the matter configuration i.e. $P(R) = 0$, which yields
\begin{eqnarray}
112 A (2 \pi (1 + 17 C R^2 + 66 C^2 R^4) +
    5 C R^2 (2 + 11 C R^2) \chi) + 5 \xi_2(R) (1 + 6 C R^2)^{4/3}=0~~~~~~\label{mc2}
\end{eqnarray}

After solving (\ref{mc})-(\ref{mc2}) simultaneously, we obtain the expressions  of  $A$ and $B$  as the functions of  $C$, the radius $R $ and mass $M$ as

\[A=\]\begin{equation} \frac{5 (1 + 6 C R^2)^{4/3} \left[10 \pi (8 C^3 R^6 + 42 C^2 R^4 + 165 C R^2 - 19) +
   3 (8 C^3 R^6 + 45 C^2 R^4 + 206 C R^2 - 56) \chi\right]}{112 \left[2 \pi  \left(66 C^2 R^4+17 C R^2+1\right)+5 C R^2 \chi  \left(11 C R^2+2\right)\right]},
\end{equation}\begin{equation}
B = \frac{R-2 M}{R \left(1+C R^2\right)^5}.
\end{equation}
It is noted that the $A$ depends on the coupling constant $\chi$ whereas $B$ does not. The numerical values of  $A$ and $B$ for ten well-known compact stars are given in Table-\ref{tab1} corresponding to a specific value of constant $C$ = 0.0006 $/km^2$.

%%%%%%%%%%%%%%%%%%%%%%%%%%%%%%%%%%%%%%%%%%%%%%%%%%%%%%%%%%%%%%%%%%%%%
\begin{figure}[!htbp]
\includegraphics[width=8.5cm]{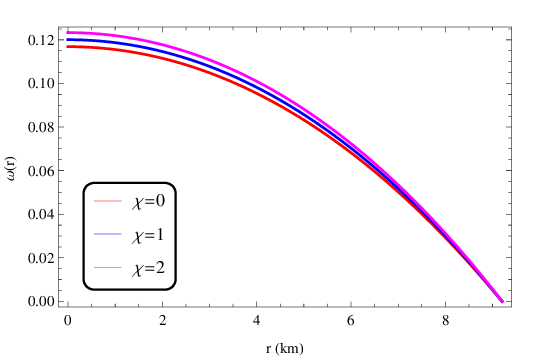}
\includegraphics[width=8.5cm]{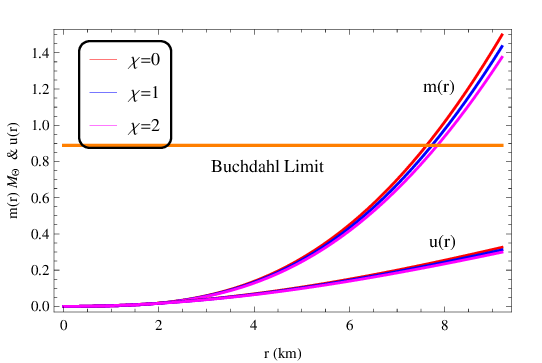}
\caption{  Profiles of  EoS parameter (Left) and mass, compactness parameter (Right) against the radial coordinate $r$ for the compact star {\it Cen X}-3 corresponding to the numerical values of constants given in Table-{\ref{tab1}}.}\label{fig4}
\end{figure}
%%%%%%%%%%%%%%%%%%%%%%%%%%%%%%%%%%%%%%%%%%%%%%%%%%%%%%%%%%

\section{ Some Relevant Physical attributes of the Solutions }\label{Sec5}

In this section, we are going to discuss some relevant attributes of the present model to make sure of its physical acceptance for presenting isotropic mass distributions in the context of $f(R, T)$ gravity.

\subsection{ Central Values of Energy Density and Pressure}

 The central values of energy density and pressure for the present solutions are obtained as

\begin{eqnarray}
\rho_c &=& \rho(0) = \frac{C \left[3 \pi(645 - 112A) +(785 - 112A) \chi \right]}{112 \left( \chi +2 \pi\right)\left(\chi +4 \pi\right)}, \label{rhc}\\
P_c &=& P(0) =  \frac{C\left[\pi(475 + 112A) + 420\chi\right]}{112  \left( \chi +2 \pi\right)\left(\chi +4 \pi\right)}. \label{pc}
\end{eqnarray}
From the above equations, we can see that the central matter density and pressure are regular in nature provided $\left( \chi +2 \pi\right)\left(\chi +4 \pi\right) \neq$ 0, this will yield the same range for the coupling constant $\chi$. Next, we shall determine the range for $\chi$.

\subsubsection{ Case-I : $\left( \chi +2 \pi\right) >$ 0}

First, let us assume $ \left( \chi +2 \pi\right) >$ 0 i.e. $ \chi > -2\pi $ . Now, $P_c > 0$ implies $ \chi > - \pi(475 + 112A)/420$. Therefore,

\begin{eqnarray}
    \chi >   {Max}\left\{-2\pi,  - \frac{\pi(475 + 112A)}{420} \right\},\label{chi1}
\end{eqnarray}

Also, $\rho_c - P_c >$ 0 gives $(\chi+4\pi)(365-112A) >$ 0 i.e. $365-112A >$ 0. Again, $\rho_c > 0$  implies $\chi (785 - 112A) > - 3 \pi(645 - 112A)$, therefore $\chi  > - 3 \pi(645 - 112A)/(785 - 112A)$ as $365-112A >$ 0.

For the value $A < 365/112$, the above inequality can be written as \\

\begin{eqnarray}
    \chi >  - 3\pi,\label{chi2}
\end{eqnarray}

Therefore, Eqs. (\ref{chi1}) and (\ref{chi2}) simultaneously give

\begin{eqnarray}
    \chi >   {Max}\left\{-2\pi,  - \frac{\pi(475 + 112A)}{420} \right\}.\label{chi3}
\end{eqnarray}

\subsubsection{ Case-II : $\left( \chi +4 \pi\right) <$ 0}

In this case, we consider $\left( \chi +4 \pi\right) <$ 0 which gives, $\left( \chi +2 \pi\right) <$ 0 i.e. $\chi  < - 2 \pi$. Now,  $\rho_c - P_c >$ 0 yields $(\chi+4\pi)(365-112A) >$ 0 which  implies, $365-112A <$0. Also, $P_c > 0$ implies $ \chi > - \pi(475 + 112A)/420$. \\
Therefore, for the value of $A > 365/112$, we obtain the range of the coupling constant $\chi$ as

\begin{eqnarray}
    - \frac{\pi}{420}(475 + 112A) < \chi < -2\pi.\label{chi4}
\end{eqnarray}

\subsubsection{ Case-III : $-4\pi < \chi < -2\pi$ }

In this case, we consider $-4\pi < \chi < -2\pi$, and for this consideration,  $\rho_c - P_c >$ 0 yields $(\chi+4\pi)(365-112A) >$ 0 which  implies, $365-112A >$0 and $P_c > 0$ implies  $\chi < - \pi(475 + 112A)/420$. Therefore, under the values of $A < 365/112$, we obtain the following range of $\chi$

\begin{eqnarray}
    -4\pi < \chi <   {Min}\left\{-2\pi,  - \frac{\pi(475 + 112A)}{420} \right\}.\label{chi5}
\end{eqnarray}

\subsection{ Maximality Criteria for Energy Density and Pressure }
To maintain the maximum values at the center with monotonically decreasing behavior, the gradients of density and pressure $\rho'(r)$ and $P'(r)$ must be zero at the center and thereafter become negative with $\rho''(0) <$ 0  and $P''(0) <$ 0. For our solutions, we obtain the density and pressure gradients as

\[\rho'(r) =\] \begin{equation} \frac{5 C^2 r \left[\lambda_1(r)\lambda_0 + 112 A \{\pi (5 + 39 C r^2 + 66 C^2 r^4 - 88 C^3 r^6) + (2 +
         15 C r^2 + 17 C^2 r^4 - 66 C^3 r^6) \chi\}\right]}{56 (1 + C r^2)^5 (1 + 6 C r^2)^{7/3} (8 \pi^2 + 6 \pi \chi+ \chi^2)}, \label{rhc}\end{equation}
      \begin{equation}
P'(r)  = \frac{5 C^2 r \left[\lambda_2(r)\lambda_0 -112 A (\pi (264 C^3 r^6 + 62 C^2 r^4 - 3 C r^2 -
         1) + (110 C^3 r^6 + 15 C^2 r^4 - 6 C r^2 - 1) \chi)\right]}{56 (1 + C r^2)^5 (1 + 6 C r^2)^{7/3} (8 \pi^2 + 6 \pi \chi+ \chi^2)}. \label{pc}\end{equation}

where
\begin{eqnarray}   \lambda_0 =  (1 + 6 C r^2)^{7/3} \nonumber
     \\
\lambda_1(r) &=& 3 \pi (57 C r^2 - 515 - 36 C^2 r^4 - 8 C^3 r^6) + (321 C r^2 - 730 - 3 C^2 r^4 - 4 C^3 r^6) \chi\nonumber
     \\
\lambda_2(r) &=& 5\pi(8 C^3 r^6 + 60 C^2 r^4+ 411 C r^2-241    ) +
  3 ( 4 C^3 r^6+ 33 C^2 r^4+ 264 C r^2-215   ) \chi \nonumber
\end{eqnarray}
\begin{equation}\end{equation}

%%%%%%%%%%%%%%%%%%%%%%%%%%%%%%%%%%%%%%%%%%%%%%%%%%%%%%%%%%%%%%%%%%%%%
\begin{figure}[!htbp]
\begin{center}
\begin{tabular}{rl}
\includegraphics[width=8.5cm]{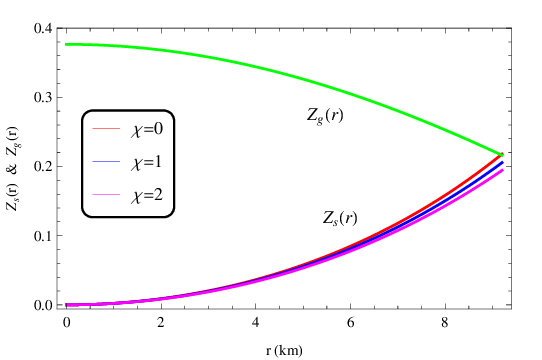}
\includegraphics[width=8.5cm]{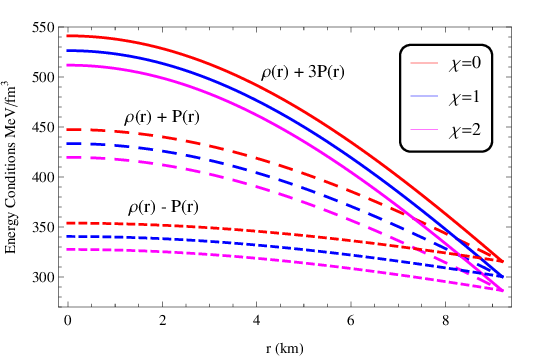}
\\
\end{tabular}
\end{center}
\caption{  Profiles of gravitational, surface redshifts (Left), and energy conditions (Right) against the radial coordinate $r$ for the compact star {\it Cen X}-3 corresponding to the numerical values of constants given in Table-{\ref{tab1}}.}\label{fig5}
\end{figure}
%%%%%%%%%%%%%%%%%%%%%%%%%%%%%%%%%%%%%%%%%%%%%%%%%%%%%%%%%%

From the above equations, we can check that $\rho'(0) = P'(0)$ = 0, and, Fig. \ref{fig2} (Right) indicates that $\rho'(r) $ and $ P'(r)$ both are negative for $0< r \leq R$.

Also, we find
\begin{eqnarray}
\rho''(0) &=& \frac{5 C^2 \left[(560 A - 1545) \pi  -(730 - 224 A) \chi\right]}{56 (8 \pi^2 + 6 \pi \chi + \chi^2)},
    \\
  P''(0) &=& \frac{5 C^2 \left[(112 A - 1545) \pi - (645 - 112 A) \chi\right]}{56 \left(\chi ^2+6 \pi  \chi +8 \pi ^2\right)}.
\end{eqnarray}

Therefore, for the star {\it Cen X}-3, $\rho''(0) \in $ [-2.3834, -2.4819] and $P''(0) \in $ [-1.3425, -1.3376] whenever $\chi \in $ [0, 2], and hence, these results guarantee that the energy density and pressure attain their maximum values at the center of the compact star, after that they are decreasing. The radially symmetric profiles of energy density (Upper panel) and pressure (Lower Panel) are displayed in Fig. \ref{fig3}, indicating that the core of the star is more dense with more pressure than the surface. Further, for physical matter distribution the EoS parameters $\omega (r) \in$ (0, 1), this is satisfied by our solutions (See Fig. \ref{fig4} (Left)).

\subsection{ Mass, Compactness Parameter, Surface Redshift, and Gravitational Redshift}
Here, we calculate the mass, compactness parameter, surface redshift, and gravitational redshift for the present model.

The mass of a matter configuration  in $f(R, T)$ gravity is obtained as
\begin{eqnarray}
m(r)&=&4\pi \int_0^r r^{*2} \rho(r^{*}) dr^{*}\nonumber
\\
&=&  [A\pi r \chi\beta_1(r) +
   C\pi \beta_2(r)  \{\sqrt{C} r (75 (1 + 6 C r^2)^{1/3} \beta_3(r)\nonumber
  \\&&-112 A (\chi - 177 C^2 r^4 \chi - 54 C^3 r^6 \chi + 2 C r^2 (240 \pi 
  + 79 \chi))  )\nonumber
  \\&&- 7125 (1 + C r^2)^3 (1 + 6 C r^2)^{1/3} \chi ArcTan[\sqrt{C} r] \} ] [13440 (8 \pi^2 + 6 \pi \chi + \chi^2) ]^{-1},
~~~~~~~~\end{eqnarray}

where

\[ \beta_1(r)=   112  AppellF1[1/2, 1/3, 1, 3/2, -6 C r^2, -C r^2] -
  672  C r^2  AppellF1[3/2, 1/3, 1, 5/2, -6 C r^2, -C r^2],\]
\begin{eqnarray}
 \beta_2(r) &=& \left[C^{3/2} (1 + C r^2)^3 (1 + 6 C r^2)^{1/3}\right]^{-1},\nonumber
 \\
  \beta_3(r) &=&95 \chi + 128 C^3 r^6 (6 \pi + \chi) +
  3 C^2 r^4 (832 \pi + 187 \chi) +
  8 C r^2 (516 \pi + 241 \chi).~~~~~~~~~
\end{eqnarray}
Here, $Appell F1[a; b_1, b_2; c; x, y]$ is the Appell hypergeometric function of two variables.

The  compactness parameter is given in terms of the mass function as
\begin{eqnarray}
u(r)&=&\frac{2m(r)}{r}.\nonumber
\\
&=& [2A\pi  \chi\beta_1 +
   2C\pi \beta_2  \{\sqrt{C}  (75 (1 + 6 C r^2)^{1/3} \beta_3-112 A (\chi - 177 C^2 r^4 \chi - 54 C^3 r^6 \chi + \nonumber
  \\
  &&2 C r^2 (240 \pi + 79 \chi)) )- 14250 (1 + C r^2)^3 (1 + 6 C r^2)^{1/3} \chi ArcTan[\sqrt{C} r]/r \}  ] \nonumber
  \\
  && \times [13440 (8 \pi^2 + 6 \pi \chi + \chi^2) ]^{-1}
\end{eqnarray}.

%%%%%%%%%%%%%%%%%%%%%%%%%%%%%%%%%%%%%%%%%%%%%%%%%%%%%%%%%%%%%%%%%%%%%
\begin{figure}[!htbp]
\begin{center}
\begin{tabular}{rl}
\includegraphics[width=8.5cm]{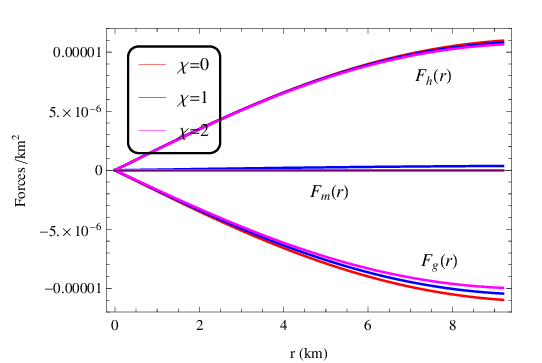}
\includegraphics[width=8.5cm]{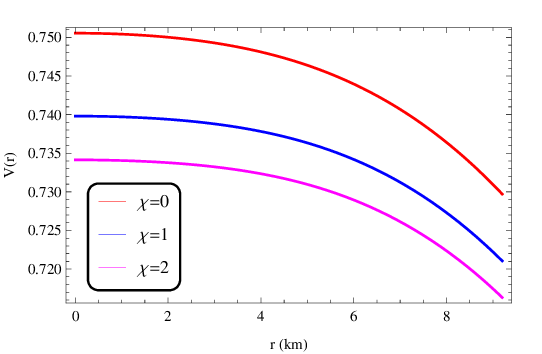}
\end{tabular}
\end{center}
\caption{  Profiles of three forces (Left) and velocity of sound (Right) against the radial coordinate $r$ for the compact star {\it Cen X}-3 corresponding to the numerical values of constants given in Table-{\ref{tab1}}.}\label{fig6}
\end{figure}
%%%%%%%%%%%%%%%%%%%%%%%%%%%%%%%%%%%%%%%%%%%%%%%%%%%%%%%%%%

Also, the surface redshift and gravitational redshift are obtained as

\[
Z_s(r)=\left[1-u(r)\right]^{-1/2}-1 = \] $ [1- [2A\pi  \chi\beta_1 +
   2C\pi \beta_2  \{\sqrt{C}  (75 (1 + 6 C r^2)^{1/3} \beta_3   
 -  112 A (\chi - 177 C^2 r^4 \chi - 54 C^3 r^6 \chi + 2 C r^2 (240 \pi + 79 \chi))  )-14250 (1 + C r^2)^3 (1 + 6 C r^2)^{1/3} \chi ArcTan[\sqrt{C} r]/r   \} ] $ \begin{eqnarray} &&\times[13440 (8 \pi^2 + 6 \pi \chi + \chi^2) ]^{-1} ]^{-1/2}-1,\end{eqnarray}
 
\begin{eqnarray}
Z_g(r)&=&e^{-\nu(r)/2}-1 = \left[B (1 + C r^2)^5\right]^{-1/2}-1.
\end{eqnarray}

The behaviors of the mass function and compactness parameter are shown in Fig. \ref{fig4} (Right), both are regular at the center of the star, also, they are finitely monotonically increasing and reached their maximum values at the surface of the star. Moreover, compactness parameter $u(r)$ is under the Buchdalh limit i.e. $u(r) < 8/9$ \cite{ha59}, indicating the physically viable celestial compact star model. The surface redshift $Z_s(r)$ is finite and monotonically increasing in nature whereas the gravitational redshift $Z_g(r)$ is finite and monotonically decreasing in nature within the interior and they have matched at the surface of the star only for $\chi =0$ as $Z_g(r)$ is not depending on $\chi$ (See Fig. \ref{fig5} (Left)). According to Buchdahl\cite{ha59} and Straumann\cite{ns84a},  the surface value of the surface redshift  $Z(R) = Z_s(R) <$ 2 for the isotropic compact star,  from Table-\ref{tab2} we can see the $ {Z(R)}$ is under the given limit for different ten well-known compact stars along with {\it Cen X-} 3 i.e. the present model is well-fitted to present the isotropic stars.

Further, the effective mass, compactness parameter, and surface redshift are obtained as

\[
m^{eff}(r)=4\pi \int_0^r r^{*2} \rho^{eff}(r^{*}) dr^{*}\]
\[=~ \frac{C r^3 \left[15 (1 + 6 C r^2)^{1/3} (43 + 26 C r^2 + 8 C^2 r^4) - 112 A\right]}{224 (1 + C r^2)^3 (1 + 6 C r^2)^{1/3}},\]
\begin{equation} \end{equation}

\[
u^{eff}(r)=\frac{2m^{eff}(r)}{r}=   \frac{C r^2 \left[15 (1 + 6 C r^2)^{1/3} (43 + 26 C r^2 + 8 C^2 r^4) - 112 A\right]}{112 (1 + C r^2)^3 (1 + 6 C r^2)^{1/3}},\]
\begin{equation} \end{equation}
\[Z_s^{eff}(r)=\left[1-u^{eff}(r)\right]^{-1/2}-1=\left[1-\frac{C r^2 \left[15 (1 + 6 C r^2)^{1/3} (43 + 26 C r^2 + 8 C^2 r^4) - 112 A\right]}{112 (1 + C r^2)^3 (1 + 6 C r^2)^{1/3}}\right]^{-1/2}-1.\]
\begin{equation} \end{equation}

%%%%%%%%%%%%%%%%%%%%%%%%%%%%%%%%%%%%%%%%%%%%%%%%%%%%
  
\begin{table*}[thth]
\tiny
\caption{ Numerical values of masses $M$, radii $R$, constants $A$, and $B$ for ten well-known compact stars corresponding to $C$ = 0.0006 $/km^2$.}
\label{tab1}       % Give a unique label
\centering
\begin{tabular}{| *{10}{c|}} 
\hline
 Compact & $M~ (M_{\odot})$ &   $R ~(km)$   &  Refs.   &  $M ~ (M_{\odot})$ &   $R ~(km)$
                    & \multicolumn{3}{c|}{A}
                            & B \\
                            \cline{7-9}
 Stars   & Observed           &   Observed         &         &     Estimated         & Estimated  & $\chi = 0$  & $\chi = 1$ & $\chi = 2$ &  \\
    \hline
Cen X-3     & 1.49 $\pm$ 0.08 &  9.178 $\pm$ 0.13 &  \cite{ml11}   &  1.49  & 9.20 & -1.6449 & -2.2112 & -2.7252 & 0.52775\\
    \hline
Her X-1     &  0.85 $\pm$ 0.15 &  8.1 $\pm$ 0.41 &  \cite{mk08}   & 0.85   & 8.50 &  -1.9211 & -2.5371  & -3.1012 & 0.64705\\
\hline
Vela X-1     & 1.77 $\pm$ 0.08 &  9.560 $\pm$ 0.08 &  \cite{ml11}   & 1.77   & 9.56 &  -1.5039 & -2.0463  & -2.5364 & 0.48218\\
\hline
LMC X-4      & 1.04 $\pm$ 0.09 &  8.301 $\pm$ 0.2 &  \cite{ml11}   & 1.04   & 8.30  & -2.0003 & -2.6312  & -3.2105 & 0.61202 \\
\hline
EXO 1785-248 & 1.30 $\pm$ 0.02 &  8.849 $\pm$ 0.4 &  \cite{f009}   &  1.30  & 8.80 &
-1.8025 &  -2.3967    & -2.9388 & 0.56142\\
\hline
4U 1538-52 & 0.87 $\pm$ 0.07 &  7.866 $\pm$ 0.21 &  \cite{ml11}   &  0.87  & 7.80 &
-2.1982 &  -2.8677  & -3.4867 & 0.64942\\
\hline
PSR J1614-2230 & 1.97 $\pm$ 0.04 &  9.69 $\pm$ 0.2 &  \cite{pd10}   &  1.97  & 9.70 & -1.4493 &  -1.9827  & -2.4638 & 0.45123 \\
\hline
PSR J1903+327 & 1.667 $\pm$ 0.021 &  9.438 $\pm$ 0.03 &  \cite{pc11}   &  1.67  & 9.40 & -1.5665 &  -2.1193  & -2.6199 &  0.49793\\
\hline
4U 1820-30 & 1.58 $\pm$ 0.06 &  9.316 $\pm$ 0.086 &  \cite{tg10}   &  1.58  & 9.60 &
-1.4883 &  -2.0281 & -2.5156 & 0.51256\\
\hline
SMC X-4 & 1.29 $\pm$ 0.05 &  8.831 $\pm$ 0.09 &  \cite{ml11}   &  1.29  & 8.80 &
-1.8025 &  -2.3967 & -2.9388 &  0.56323\\
\hline
\end{tabular}\label{table3}
\end{table*}
%%%%%%%%%%%%%%%%%%%%%%%%%%%%%%%%%%%%%%%%%%%%%%%%%%%%%%%%%%%%%%%%%%%%%%%%%%

\subsection{  Energy Conditions }

 It is expected that every solution of the Einstein field equations for presenting physical matter objects needs to satisfy all the energy conditions whether in modified gravity or in Einstein gravity because it tells the presence of ordinary and exotic matter within the matter objects. All the energy conditions are defined as some inequalities depending
 on the energy-momentum tensor $T_{\mu\nu}$. The  Null energy condition (NEC), Weak energy condition (WEC),  and Strong energy condition (SEC) are defined as $T_{\mu\nu} k^\mu k^\nu \geq$  0,  $T_{\mu\nu} U^\mu U^\nu \geq$  0, and $\left(T_{\mu\nu} -\frac{1}{2} T g_{\mu\nu}\right)U^\mu U^\nu\geq$  0, respectively,  where $k^\nu$ is a null vector and $U^\mu$ is a timelike vector. Moreover, the  Dominant energy condition (DEC) is defined as $T_{\mu\nu} U^\mu U^\nu \geq$  0, where $U^\nu$ is a timelike vector but $T_{\mu\nu}U^\nu$ not spacelike.  Therefore, for the given diagonal energy-momentum tensor all these energy conditions read as\cite{step,wald}

\begin{eqnarray}
 {NEC} &:&  \rho(r)+P(r) \geq  0,
\\
 {WEC} &:& \rho(r) \geq  0,~\rho(r)+P(r) \ge 0,
\\
 {SEC} &:&  \rho(r)+3P(r) \ge 0,
\\
 {DEC} &:&  \rho(r) \geq  0,~\rho(r)-P(r) \ge 0.\label{EC}
\end{eqnarray}

 Figs.\ref{fig1} (Right) and \ref{fig5} (Right) ensure that the present solutions are nicely met all the energy conditions. Consequently,  our solutions represent physical matter objects.

\section{ EQUILIBRIUM Via TOV Equation}\label{Sec6}
The equilibrium i.e. the dynamical balance of any celestial object takes place under the simultaneous action of different internal forces and this situation can be described by the generalized {\it Tolman-Oppenheimer-Volkoff} (TOV) equation. The generalized TOV equation for an isotropic matter distribution in the framework of $f(R, T)$ gravity is given as\cite{tm19}

  \begin{eqnarray}
&&-\frac{\nu'(r)}{2}\left[\rho(r)+P(r)\right]-\frac{dP(r)}{dr}+\frac{\chi}{8\pi+2\chi}\left[P'(r)-\rho'(r)\right]=0,\label{F}
\end{eqnarray}
The above equation can be written as

 \begin{eqnarray}
F_g(r)+F_h(r)+F_m(r)=0.
\end{eqnarray}

where, $F_g(r) = -\frac{\nu'(r)}{2}[\rho(r)+p_r(r)]$ termed as the gravitational force,  $F_h(r) = -\frac{dp_r(r)}{dr}$ termed as the hydrostatic force, and $F_m(r) = \frac{\chi}{8\pi+2\chi}\left[P'(r)-\rho'(r)\right]$ termed as the additional force that occurred inside the matter configuration due to the coupling between the matter and geometry in $f(R, T)$ gravity. It is noted that for the dynamical equilibrium in $f(R, T) = R + \chi T$ gravity couple constant $\chi \neq -4\pi$. However, for $\chi = 0$, Eq. (\ref{F}) becomes

 \begin{eqnarray}
&&-\frac{\nu'(r)}{2}\left[\rho(r)+P(r)\right]-\frac{dP(r)}{dr}=0,\label{F1}
\end{eqnarray}
This is the TOV equation for isotropic matter objects in standard general relativity, which is the same as the result of Oppenheimer and Volkoff\cite{jr39}.

%%%%%%%%%%%%%%%%%%%%%%%%%%%%%%%%%%%%%%%%%%%%%%%%%%%%%%%%%%%%%%%%%%%%%
\begin{figure}[!htbp]
\begin{center}
\begin{tabular}{rl}
\includegraphics[width=8.5cm]{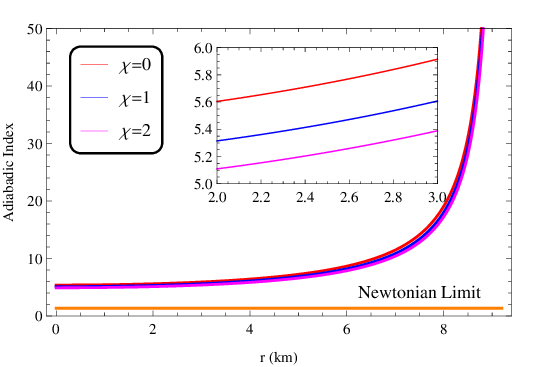}
\includegraphics[width=8.5cm]{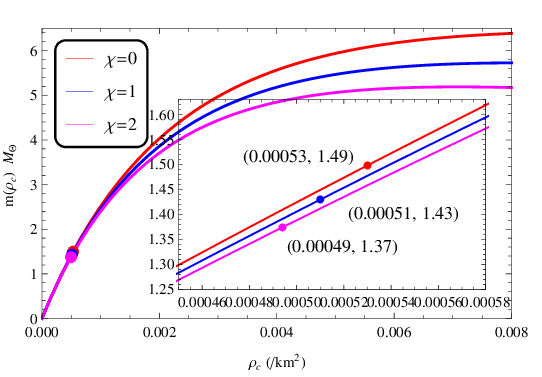}
\\
\end{tabular}
\end{center}
\caption{ Profiles of adiabatic index against the radial coordinate $r$ (Left) and mass  against the central density $\rho_c$ for the compact star {\it Cen X}-3 corresponding to the numerical values of constants given in Table-{\ref{tab1}}.}\label{fig7}
\end{figure}
%%%%%%%%%%%%%%%%%%%%%%%%%%%%%%%%%%%%%%%%%%%%%%%%%%%%%%%%%%

For our reported solutions, we obtain the following expressions for the forces

\[
 F_g(r) =\] \[\frac{5 C^2 r \left[112 A (1 - 3 C r^2 - 44 C^2 r^4) -
   5 (1 + 6 C r^2)^{1/3} (241 + 1035 C r^2 - 2526 C^2 r^4 - 368 C^3 r^6 - 48 C^4 r^8)\right]}{112 (4\pi+ \chi)(1 +Cr^2)^5 (1 + 6Cr^2)^{4/3}},\]
   \begin{equation} \end{equation}
\[
 F_h(r)=\] \[\frac{5 rC^2  [112 A (\pi (264 C^3 r^6 + 62 C^2 r^4 - 3 C r^2 -1) + (110 C^3 r^6 + 15 C^2 r^4 - 6 C r^2 - 1) \chi)-\lambda_2(r) \lambda_0 ]}{56 (1 + C r^2)^5 (1 + 6 C r^2)^{7/3} (8 \pi^2 + 6 \pi \chi+ \chi^2)},\]
 \begin{equation} \end{equation}
\[ F_m(r) =\]\[ \frac{5\chi C^2 r \left[(1 + 6 C r^2)^{7/3} (85 + 471 C r^2 + 102 C^2 r^4 + 16 C^3 r^6) -112 a (1 + 9 C r^2 + 32 C^2 r^4 + 44 C^3 r^6)\right] }{112(2 \pi + \chi) (4 \pi + \chi) (1 + C r^2)^5 (1 + 6 C r^2)^{7/3} }.
\]\begin{equation} \end{equation}

The interplay of the above three forces is shown graphically in Fig. \ref{fig6} (Left), which ensures that $F_g(r)+F_h(r)+F_m(r)$ = 0 inside the star with $F_g(r)$ as an attractive force acts in inwards direction, $F_h(r)$ as a repulsive force acts in outward direction and the coupling force $F_m(r)$ having a very negligible effect on achieving the equilibrium position. Consequently,  the present solutions are able to represent equilibrium matter configurations in $f(R, T)$ gravity.

\section{ STABILITY ANALYSIS}\label{Sec7}
In this section, we are willing to investigate the stability of the preset model via {\bf (i)} Causality condition, {\bf (ii)} Adiabatic index, and {\bf (iii)} {\it Harrison-Zeldovich-Novikov's} static stability condition.

%%%%%%%%%%%%%%%%%%%%%%%%%%%%%%%%%%%%%%%%%%%%%%%%%%%%%%%%%%%%%%%%%%%%%
\begin{figure}[!htbp]
\begin{center}
\begin{tabular}{rl}
\includegraphics[width=8.4cm]{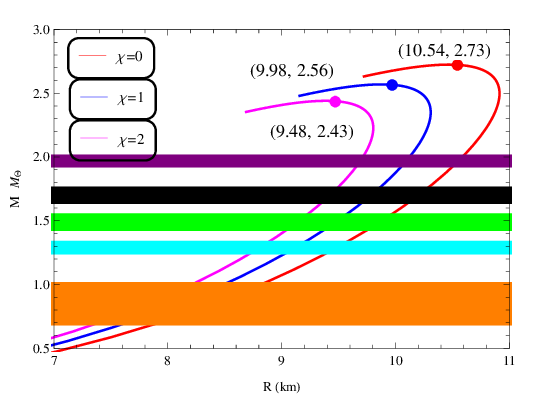}
\includegraphics[width=8.4cm]{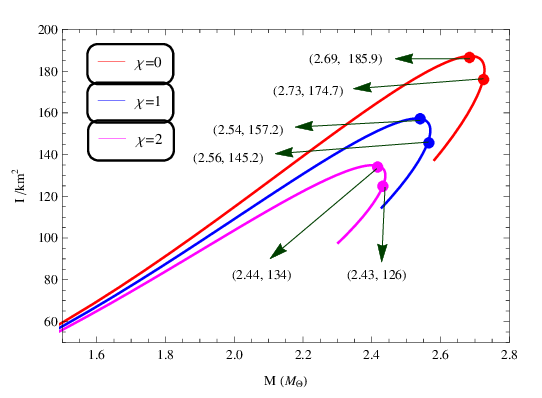}
\\
\end{tabular}
\end{center}
\caption{ Profiles mass $M$ against the surface radius $R$ (Left) and Moment of Inertia $I$  against the mass $M$ (Right) for the compact star {\it Cen S}-3 corresponding to the numerical values of constant given in Table-{\ref{tab1}}. The   strips in $M-R$ are as follows: 1. Orange : {\it Her X}-1 , 2. Cyan : {\it SMC X}-4, 3. Green : {\it Cen X}-3, 4. Black : {\it Vela X}-1, 5. Purple : {\it PSR J1614-2230}.}\label{fig8}
\end{figure}

%%%%%%%%%%%%%%%%%%%%%%%%%%%%%%%%%%%%%%%%%%%%%%%%%%%%%%%%%%

\subsection{ Causality Condition}
According to the causality condition, the velocity of sound $V(r)$ within the stable compact star is always less than the velocity of light $c$ i.e. $0 \leq V(r) < 1$ (In the unit $c = 1$), otherwise it will become an unstable situation. the velocity of sound $V(r)$ is defined as

\[
V^2(r)=\left[{dP_r(r) \over d\rho(r)}\right] \]
\[ \frac{\lambda_0\delta_1(r)-112 A \left(\chi  \left(1+6 C r^2-15 C^2 r^4-110 C^3 r^6\right)+\pi  \left(1+3 C r^2-62 C^2 r^4-264 C^3 r^6\right)\right)}{\lambda_0\delta_2(r)  -112 A (\pi (5 + 39 C r^2 + 66 C^2 r^4 - 88 C^3 r^6) + (2 + 15 C r^2 + 17 C^2 r^4 - 66 C^3 r^6) \chi)},
\]
 \begin{equation} \end{equation}
where
\begin{eqnarray}
\delta_1(r) &=&5 \pi (241 - 411 C r^2 - 60 C^2 r^4 - 8 C^3 r^6) + 3 (215 - 264 C r^2 - 33 C^2 r^4 - 4 C^3 r^6)\chi,\nonumber
\\
\delta_2(r) &=& 3 \pi (515 - 57 C r^2 + 36 C^2 r^4 + 8 C^3 r^6) + (730 - 321 C r^2 + 3 C^2 r^4 + 4 C^3 r^6) \chi.
\end{eqnarray}
The exact behavior of $V(r)$ is shown in Fig. \ref{fig6} (Right) for our model compact star {\it Cen} X-3, clearing that the solutions satisfy the causality condition. Therefore, it ensures stable matter distribution.

\subsection{Adiabatic Index}
 The adiabatic index plays a very important role in the thermal stability of the stellar model.  Chandrasekhar\cite{sc64} first developed the idea of stability against the very small radial adiabatic disturbance within the stellar object, and later,  several researchers elaborated the study of stability condition\cite{hb64, hh75, wh76, chc17, rc93, mm13}. According to their results, the adiabatic index $\Gamma(r)$ should be bigger than 4/3, Newtonian limit, within the stellar object. The adiabatic index $\Gamma(r)$ is defined as

\begin{eqnarray}
\Gamma(r)= \left(1+\frac{\rho(r)}{P(r)}\right) \{V(r)\}^2.
\end{eqnarray}

  Fig. \ref{fig7} (Left) indicates that  $\Gamma(r) > 4/3$ for our proposed solutions  in $f(R, T)$ gravity. Consequently, the present stellar model is stable with respect to the adiabatic index.

%%%%%%%%%%%%%%%%%%%%%%%%%%%%%%%%%%%%%%%%%%%%%%%%%%%%%%%%%%%
\begin{table*}[thth]
\tiny
 %able caption is above the table
\caption{Numerical values of central density, surface density, central pressure, central EoS parameter, surface redshift, and  compactness parameter at the surface $r = R$ for ten well-known compact stars corresponding to the numerical values of constants given in Table-\ref{tab1}}.\label{tab2}
\centering
\begin{tabular}{| *{9}{c|}}
\hline
 \multicolumn{8}{|c|}{For $\chi$ = 0}\\
\hline
Compact   &  $\rho_c ~ (10^{14})$ & $\rho(R) ~(10^{14})$ & $P_c ~(10^{34})$& $\omega_c = P_c/\rho_c$  &  $Z(R)$ & $ u(R)$ & Buchdahl \\
Stars & $gm/cm^3$ & $gm/cm^3$ & $gm/cm^2$ &  & & & Limit\cite{ha59}\\
\hline
Cen X-3  &   7.1428     & 5.6275  & 7.5034 & 0.116884 & 0.217907 & 0.325826 & $< 8/9$\\
\hline
Her X-1     &  7.4092   &  6.0081 & 6.7052 & 0.100694  & 0.189743 & 0.293530 & $< 8/9$\\
\hline
Vela X-1    &    7.0068 & 5.4383  & 7.9109 & 0.125625 & 0.232824 & 0.342042 & $< 8/9$\\
\hline
LMC X-4  &   7.4856     & 6.1195  & 6.4764 & 0.096265 & 0.181919 & 0.284146 & $< 8/9$\\
\hline
EXO 1785-248  & 7.2948  & 5.8431  & 7.0480 & 0.107502 & 0.201669 & 0.307484 & $< 8/9$\\
\hline
4U 1538-52   &  7.6766  & 6.4025 & 5.9043 & 0.085579 & 0.162834 & 0.260456 & $< 8/9$\\
\hline
PSR J1614-2230 & 6.9541 & 5.3659 & 8.0686 & 0.129100 & 0.238698 & 0.348268 & $< 8/9$\\
\hline
PSR J1903+327 & 7.0671  & 5.5218  & 7.7301 & 0.121705 & 0.226160 & 0.334871 & $< 8/9$\\
\hline
4U 1820-30   &  6.9917  & 5.4175  &7.9560  & 0.126613 & 0.234498 & 0.343826 & $< 8/9$\\
\hline
SMC X-4   &  7.2948     & 5.8431 & 7.0480 & 0.107502 & 0.201669 & 0.307484 & $< 8/9$\\
\hline
\end{tabular}
\\
\centering
\begin{tabular}{| *{9}{c|}}
\hline
 \multicolumn{8}{|c|}{For $\chi$ = 1}\\
\hline
Compact   &  $\rho_c ~ (10^{14})$ & $\rho(R) ~(10^{14})$ & $P_c ~(10^{34})$& $\omega_c = P_c/\rho_c$ &  $Z(R)$ & $ u(R)$ & Buchdahl \\
Stars & $gm/cm^3$ & $gm/cm^3$ & $gm/cm^2$ &  & & & Limit\cite{ha59}\\
\hline
Cen X-3  &   6.8987     & 5.3552  & 7.4449 & 0.120077 & 0.205531 & 0.311913 & $< 8/9$\\
\hline
Her X-1     &  7.1765   &  5.7443 & 6.6924 & 0.103761  & 0.180245 & 0.282114 & $< 8/9$\\
\hline
Vela X-1    &    6.7580 & 5.1630  & 7.8259 & 0.128849 & 0.218784 & 0.326797 & $< 8/9$\\
\hline
LMC X-4  &   7.2567     & 5.8589  & 6.4751 & 0.099282 & 0.173158 & 0.273414 & $< 8/9$\\
\hline
EXO 1785-248  & 7.0568  & 5.5752 & 7.0166 & 0.110633 & 0.190995 & 0.295015 & $< 8/9$\\
\hline
4U 1538-52   &  7.4585  & 6.1513 & 5.9286 & 0.088444 & 0.155755 & 0.251368 & $< 8/9$\\
\hline
PSR J1614-2230 &6.7038  & 5.0898 & 7.9727 & 0.132329 & 0.223977 & 0.332497 & $< 8/9$\\
\hline
PSR J1903+327 & 6.8203  & 5.2477  & 7.6571 & 0.124918 & 0.212875 & 0.320221 & $< 8/9$\\
\hline
4U 1820-30  &6.7425    & 5.1420  & 7.8679 & 0.129839 & 0.220266 & 0.328430 & $< 8/9$\\
\hline
SMC X-4   &  7.0568    & 5.5752 & 7.0166 & 0.110633 & 0.190995 & 0.295015 & $< 8/9$\\
\hline
\end{tabular}
\\
\centering
\begin{tabular}{| *{9}{c|}}
\hline
 \multicolumn{8}{|c|}{For $\chi$ = 2}\\
\hline
Compact   &  $\rho_c ~ (10^{14})$ & $\rho(R) ~(10^{14})$ & $P_c ~(10^{34})$ & $\omega_c = P_c/\rho_c$ &  $Z(R)$ & $ u(R)$ & Buchdahl \\
Stars & $gm/cm^3$ & $gm/cm^3$ & $gm/cm^2$ &  & & & Limit\cite{ha59}\\
\hline
Cen X-3 &   6.6604     & 5.1081  & 7.3821& 0.123323 & 0.194358 & 0.298979 & $< 8/9$\\
\hline
Her X-1   &  6.9481    &  5.5028 & 6.6711 & 0.106830  & 0.171548 & 0.271415 & $< 8/9$\\
\hline
Vela X-1   &  6.5158   & 4.9143  & 7.7393 & 0.132159 & 0.206203 & 0.31268 & $< 8/9$\\
\hline
LMC X-4  &   7.0317    & 5.6196  & 6.4644 & 0.102289 & 0.165104 & 0.263334 & $< 8/9$\\
\hline
EXO 1785-248 & 6.8238  & 5.3308  & 6.9783 & 0.113786 & 0.181280 & 0.283371 & $< 8/9$\\
\hline
4U 1538-52  &  7.2432  & 5.9191 & 5.9419 & 0.091277 & 0.149183 & 0.242780 & $< 8/9$\\
\hline
PSR J1614-2230 &6.4603 & 4.8406 & 7.8766 & 0.135659 & 0.210824 & 0.317916 & $< 8/9$\\
\hline
PSR J1903+327 &6.5798 & 4.9996  & 7.5813 & 0.128203 & 0.200932 & 0.306633 & $< 8/9$\\
\hline
4U 1820-30  &  6.4999  & 4.8931  & 7.7786 & 0.133155 & 0.207523 & 0.314181 & $< 8/9$\\
\hline
SMC X-4   &  6.8238  & 5.3308    & 6.9783 & 0.113786 & 0.18128 & 0.283371 & $< 8/9$\\
\hline
\end{tabular}
\end{table*}

%%%%%%%%%%%%%%%%%%%%%%%%%%%%%%%%%%%%%%%%%%%%%%%%%%%%%%%%%%%%%%%%%%%%%%%%%%%%%%%%%%%%%%%%%%

\subsection{Harrison-Zeldovich-Novikov's Static Stability Condition}
According to Harrison et al.\cite{har65} and Zeldovich-Novikov\cite{zel}, the mass of the stable stellar object is positively increasing against its central pressure, otherwise unstable stellar object.

The mass as a function of terms of $\rho_c$ is obtained as
\begin{eqnarray}
 M(\rho_c) &=&  [A\pi R \chi {H}_2 +
   {H}\pi  {H}_3  \{\sqrt{ {H}_1} R (75 (1 + 6  {H}_1 R^2)^{1/3}  {H}_4-\nonumber
  \\
  &&112 A (\chi - 177  {H}_1^2 R^4 \chi - 54  {H}_1^3 R^6 \chi + 2  {H}_1 R^2 \times\nonumber
  \\
  &&(240 \pi + 79 \chi)) )- 7125 (1 +  {H}_1 R^2)^3 (1 + 6 {H}_1 R^2)^{1/3} \chi ArcTan[\sqrt{ {H}_1} R] \} ]\nonumber
  \\
  && [13440 (8 \pi^2 + 6 \pi \chi + \chi^2) ]^{-1},
 \end{eqnarray}
where
\begin{eqnarray}
 {H}_1 &=& 112 \rho_c  \left(\chi ^2+6 \pi  \chi +8 \pi ^2\right)/[(785-112 A) \chi +3 \pi  (645-112 A)],\nonumber
\\
 {H}_2 &=& 112  AppellF1[1/2, 1/3, 1, 3/2, -6  {H}_1 R^2, - {H}_1 R^2] \nonumber
  \\
  &&-
  672  {H}_1 R^2  AppellF1[3/2, 1/3, 1, 5/2, -6  {H}_1 R^2, - {H}_1 R^2],\nonumber
  \\
 {H}_3 &=& \left[ {H}_1^{3/2} (1 +  {H}_1 R^2)^3 (1 + 6  {H}_1 R^2)^{1/3}\right]^{-1},\nonumber
 \\
 {H}_4 &=&95 \chi + 128  {H}_1^3 R^6 (6 \pi + \chi) +
  3  {H}_1^2 R^4 (832 \pi + 187 \chi) +
  8  {H}_1 R^2 (516 \pi + 241 \chi).~~~~~~
\end{eqnarray}

The profile of $M(\rho_c)$ is displayed in Fig. \ref{fig7} (Right), it is clear that $M(\rho_c)$  is nicely met with the required condition. Therefore, this result is also in favor of representing stable matter configurations. Moreover, the central densities of the model compact star {\it Cen X}-3 are 0.00053 $/km^2$, 0.00051 $/km^2$, and  0.00049 $/km^2$ corresponding to the coupling constant $\chi$ = 0, 1, and 2 associated with the mass 1.49 $M_\odot$, 1.43 $M_\odot$, and 1.37 $M_\odot$, respectively (See Fig. \ref{fig7} (Right)).

%%%%%%%%%%%%%%%%%%%%%%%%%%%%%%%%%%%%%%%%%%%%%%%%%%%%%%%%%%%%%%%%%%%%%
\begin{figure}[!htbp]
\begin{center}
\begin{tabular}{rl}
\includegraphics[width=8.8cm]{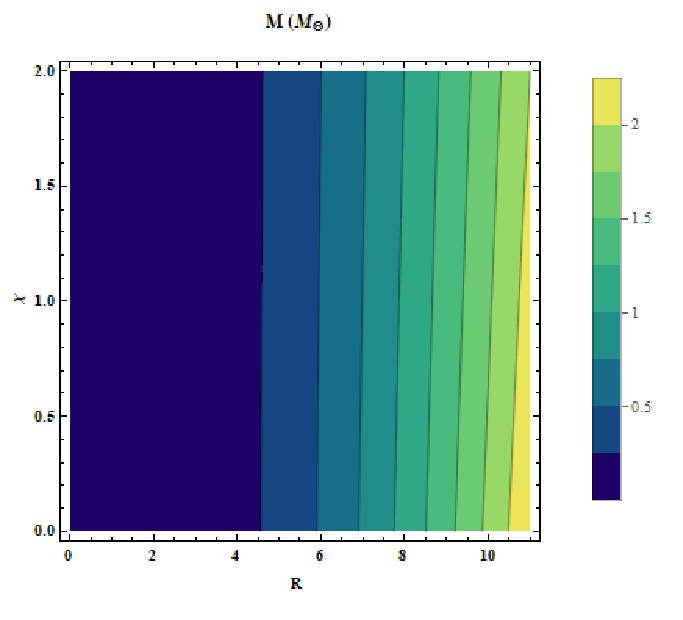}
\\
\end{tabular}
\end{center}
\caption{  The equi-mass contour plot on the $R$ ? $\chi$ plane  for the compact star {\it Cen X} -3 corresponding to the numerical values of constants given in Table-\ref{tab1}.}\label{fig9}
\end{figure}
%%%%%%%%%%%%%%%%%%%%%%%%%%%%%%%%%%%%%%%%%%%%%%%%%%%%%%%%%%

\section{Moment of inertia }\label{Sec8}
Lattimer and Prakash \cite{latt} defined the moment of inertia of a uniformly rotating stellar object as
\begin{eqnarray}
I = {8\pi \over 3} \int_0^R r^4 [\rho(r)+P(r)] e^{(\lambda(r)-\nu(r))/2} ~{\bar{\omega} \over \Omega}~dr
\end{eqnarray}

where $\Omega$ is the angular velocity of the stellar object, and $\bar{\omega}$ stands for the rotational drag satisfying the following equation
\begin{eqnarray}
{d \over dr} \left(r^4 j ~{d\bar{\omega} \over dr} \right) =-4r^3\bar{\omega}~ {dj \over dr} .
\end{eqnarray}
The above equation is known as  Hartle's equation\cite{hart} where $j=e^{-(\lambda(r)+\nu(r))/2}$ with $j(R)=1$. Now, the moment of inertia $I$ up to the maximum mass $M_{max}$ can be defined as\cite{bejg}
\begin{equation}
I = {2 \over 5}  (1+x  ) {MR^2},
\end{equation}
where $x = (M/R)\cdot km/M_\odot$.

The nature of the moment of inertia $I$ with respect to the mass is demonstrated in Fig. \ref{fig8} (Right) for the model compact star {\it Cen X}-3. One can see from this figure that the moment of inertia is increasing against the mass up to a certain range then it is decreasing. Also, $I_{max}$ and $M_{max}$ decrease whenever $\chi$ is increasing.

\section{Results and Conclusion}\label{Sec9}
In this article, we have presented a new model for the static and spherically symmetric isotropic stellar compact stars based on the {\it Durgapal-V} metric in the context of the $f(R, T) = R + \chi T$ gravity. The present solutions of the Einstein field equations are analyzed graphically for the well-known compact star {\it Cen X} -3 and numerically for compact stars {\it Cen X} -3 along with {\it Her X} -1, {\it Vela X} -1, {\it LMC X} -4, {\it EXO } 1785-248, {\it 4U } 1538-52, {\it PSR J}1614-2230, {\it PSR J}1909+327, {4U }1820-30 and {\it SMC X} -4 corresponding to the values of coupling constant $\chi \in \{0, 1, 2\}$. Interestingly, we have found the following key features of the present model:

\begin{itemize}

 \item  {\bf Metric Potentials :} The considered Durgapal-V metric potentials  $e^{\nu(r)}$ and $e^{-\lambda(r)}$ are singularity free within the interior of the star, $e^{\nu(r)}$ is finitely increasing and $e^{-\lambda(r)}$ in finitely decreasing in nature with $e^{\nu(0)} = b$ = 0.52776, $e^{\nu(R)}$ = 0.67609, these values are independent of the $f(R, T)$ gravity coupling parameter $\chi$,  and $e^{-\lambda(0)}$ = 1, $e^{-\lambda(R)}$ = 0.67609, 0.65149, 0.63089 for the compact star {\it Cen X}-3 corresponding to $\chi $ = 0, 1, 2. Moreover, $e^{\nu(r)}$ and $e^{-\lambda(r)}$ together meet with the Schwarzchild solution's metric $g_{rr}$ at the surface of the compact star for $\chi$ = 0, clear from Fig. \ref{fig1} (Left) and numerical results. It noted that  $e^{-\lambda(r)}$  only decreases whenever coupling parameter $\chi$ increases, and hence,  coupling parameter $\chi$ effects only on the values of $e^{-\lambda(r)}$ not in non-singular nature of it. Therefore, all these results ensure that the Durgapal-V metric potentials are suitable for generating a non-singular model for celestial compact stars in the framework of $f(R, T)$ gravity.

 \item {\bf Energy Density and Pressure :} The energy density $\rho(r)$, and pressure $P(r)$ both are regular, positively finite with maximum values at the center and thereafter, decreasing in nature towards the surface of the matter sphere, clear from Figs.\ref{fig1} (Right) and \ref{fig2} (Left), respectively. The decreasing nature of $\rho(r)$ and $P_r(r)$ are also confirmed from the behaviors of energy gradient and pressure gradient, both are negative in $o < r \leq R$ (See Fig. \ref{fig2} (Right)). The radially symmetric profiles of $\rho(r)$ and $P_r(r)$  are demonstrated in Fig. \ref{fig3}, which shows exactly the same decreasing behavior of energy density and pressure from the center to the surface of the compact star. Moreover, the maximum values of the energy density, $\rho(0) \in$ [400.69 $Mev/fm^3$, 373.62 $Mev/fm^3$] for $\chi \in$ [0, 2], and the minimum values, $\rho(R) \in$ [315.68 $Mev/fm^3$, 286.54 $Mev/fm^3$] for $\chi \in$ [0, 2]. Also, the maximum values of the pressure, $P(0) \in$ [46.84 $Mev/fm^3$, 46.08 $Mev/fm^3$] for $\chi \in$ [0, 2] with minimum value $P(R)$ = 0 for all $\chi \in$ [0, 2]. The present solutions represent the physical matter distributions because of its EoS parameter $\omega (r) \in$  (0, 1), clear from Fig. \ref{fig4} (Left) and the numerical values of $\omega_c$, given in Table-\ref{tab2}. To make our model more reliable, we have estimated the numerical values of central and surface densities and central pressure for ten well-known compact stars, given in Table-\ref{tab2} for $\chi$ = 0, 1, 2. One can see the central and surface densities both are of order $10^{14}~gm/cm^3$ and central pressure is of order $10^{34}~gm/cm^2$ for all these stars,  which are fine with the observational data.  It is worth mentioning that increasing $\chi$ from 0 to 2 reduces the values of $\rho(r)$ and $P(r)$ i.e. the compact stars have more energy density and pressure in the standard Einstein's gravity than modified $f(R, T)$ gravity, therefore, the modified $f(R, T)$ gravity is more suitable to support the long-term stable compact stars than the standard Einstein's gravity in Durgapal-V spacetime. \\

\item {\bf Mass, Compactness Parameter Gravitational and Surface Redshifts :} The mass function $m(r)$ and compactness parameter $u(r)$ for our reported solutions in $f(R, T)$ gravity are singularity free, positively finitely, and more interestingly, increasing in nature inside the star (See Fig. \ref{fig4} (Right)). We can see that both $m(r)$ and $u(r)$ decrease for the increasing values of $\chi$ as expected from the behaviors of $\rho(r)$ and $P(r)$, which supports the modified $f(R, T)$ gravity to hold the more stable compact stars than the Einstein gravity in Durgapal-V spacetime. Moreover, the compactness parameter $u(r)$  satisfies the  Buchdahl Limit i.e. $u(r) = 2m(r)/r < 8/9$ for $0 \leq r \leq R$ (See Fig. \ref{fig4} (Right) and Table-\ref{tab2}). The gravitational redshift $Z_g(r)$ and surface redshift $Z_s(r)$ are both finite and positive,  $Z_g(r)$ is monotonically decreasing whereas $Z_s(r)$ is monotonically increasing in nature and they have matched at the surface of the star for only $\chi = 0$ as $Z_g(r)$ is not depending on $\chi$. Furthermore, we have estimated the surface redshift at the surface of ten well-known compact stars $Z(R)$, all these values indicate that $Z(R) < 2$ (See Table-\ref{tab2}) i.e under the range provided in Refs.\cite{ha59,ns84a}.

\item  {\bf Energy Conditions :}  The energy conditions namely, NEC, WEC, SEC, and DEC act as the indicators for confirming the physical and nonphysical nature of the matter distributions. The solutions representing matter distributions are formed with the physical matter if the solutions satisfy  NEC, WEC, SEC, and DEC, otherwise nonphysical. Figs. \ref{fig1} (Left) and \ref{fig5} (Right) ensure that the present solutions nicely satisfy all the energy conditions, and therefore, our solutions support the physical matter configurations.

\item  {\bf Equilibrium :} The study of equilibrium is necessary because it describes the interplay of the interior forces of compact objects to become dynamically stable avoiding the gravitational collapse. In the context of $f(R, T)$ gravity, the isotropic matter configurations remain in an equilibrium position under the action of gravitational force $F_g(r)$, hydrostatic force $F_h(r)$, and an additional force $F_m(r)$ generated from the coupling between matter and geometry by satisfying the generalized TOV equation, one can see the generalized TOV equation satisfying result for our proposed solutions in $f(R, T)$ gravity (See Fig. \ref{fig6} (Left)). Therefore, the solutions representing matter configurations are in an equilibrium state avoiding gravitational collapse. In the equilibrium state, $F_g(r)$ and $F_h(r)$ act as the attractive force and repulsive force, respectively with a negligible amount of additional force $F_m(r)$. Moreover, $F_g(r)$ and $F_h(r)$ both are maximum at the surface of the star, and they are decreasing with increasing values of $\chi$.

\item  {\bf Stability : } We have ensured the stability of the present model with the help of the causality condition, adiabatic index, and  Harrison-Zeldovich-Novikov's static stability condition. The solutions satisfy the causality condition $0 \leq V(r) < 1$ within the interior of the star for $\chi =$ 0, 1, 2 (See Fig. \ref{fig6} (Right)), adiabatic index $\Gamma (r)>$ $\frac{4}{3}$  i.e. satisfy the Newtonian limit (See Fig. \ref{fig7} (Left)) in the interior the star {\it Cen X}-3 for the same values of $\chi$, moreover, the mass profile is increasing against the central density (See Fig. \ref{fig7} (Right)). Therefore, all these analyses indicate that our reported solutions are comfortable to hold static stable matter configuration. In addition, we can see in Fig. \ref{fig7} (Right) that mass $m$ = \{1.49 $M_\odot$, 1.43 $M_\odot$, 1.37 $M_\odot$ \} at the central density $\rho_c$ = \{0.00053 $/km^2$, 0.00051 $/km^2$, 0.00049 $/km^2$\} corresponding to $\chi$ = \{0, 1, 2\}, respectively, these results also suggest that the star is more massive with more central density in Einstein's gravity than modified $f(R, T)$ gravity.

\item  {\bf $M-R$ and $I-M$ Relations :} In the framework of $f(R, T)$ gravity, the present solutions holding maximum masses $M_{max}$ against the surface radius $R$ are shown in Fig. \ref{fig8} (Left) for $\chi \in$  \{0, 1, 2\}. One can see that the maximum masses $M_{max}$ = 2.73 $M_\odot$, 2.56 $M_\odot$, and 2.43 $M_\odot$ occurred at the surface radius $R$ = 10.54 $km$, 9.98 $km$, and 9.48 $km$ for $\chi$ = 0, 1, and 2, respectively. All these results ensure that the increasing $\chi$ affects the maximum mass to reduce it. It is worth mentioning that all the presented maximum masses under the mass limit of  Rhoades-Ruffini i.e $ M_{max} \leq 3.2 M_{\odot}$\cite{ruff}.  Further, we have demonstrated the moment of inertia $I$ against mass $M$ in Fig. \ref{fig8} (Right), the moment of inertia $I$ increases for increasing some certain values of mass $M$ then decreases i.e. the maximum values of the moment of inertia $I_{max}$ have not occurred at $M_{max}$.  Here,  we have estimated that $I_{max} = 185.9 ~km^{-2},~ 157.2~km^{-2},~134 ~km^{-2}$ occurred at $M$ = 2.69 $M_{\odot}$, 2.54 $M_{\odot}$, 2.44 $M_{\odot}$ corresponding to $\chi $ = 0, 1, 2, respectively. However, at the $M_{max} = 2.73~ M_\odot,~ 2.56 ~M_\odot, ~2.43~M_\odot$  the values of moment of inertia $I = 174.7 ~km^{-2},~ 145.2~km^{-2},~126 ~km^{-2}$ for $\chi $ = 0, 1, 2, respectively. The maximum mass profile is also demonstrated in the $R$-$\chi$ plan in Fig. \ref{fig9}, which also clear that maximum mass decreases with increasing values of $\chi$, and hence, this result also in favor of more stable mass distribution in modified $f(R, T)$ gravity than the Einstein gravity.

\end{itemize}

 Finally, we can say that all the significant results obtained from the different graphical and numerical analyses ensure that the present proposed model for isotropic celestial compact stars is physically well-behaved, stable, and staying in an equilibrium position in the Durgapal-V spacetime under the framework of $f(R, T)$ gravity. In this connection, the modified $f(R, T)$ gravity is more capable of holding the long-term stable isotropic stars than the standard Einsten's gravity in the  Durgapal-V spacetime. Therefore, the scientific community may be inspired by this present work for doing fruitful research work in the Durgapal-V spacetime under other different modified theories of gravity in the future.

\subsection*{Acknowledgments}
Farook Rahaman would like to thank the authorities of the Inter-University Centre for Astronomy and Astrophysics, Pune, India for providing the research facilities.

\end{document}